\documentclass[%
 reprint,
 amsmath,amssymb,
 aps,
]{revtex4-2}
\usepackage{graphicx}
\usepackage{dcolumn}
\usepackage{bm}
\usepackage{xcolor}
\usepackage{multirow}
\usepackage{tabularx}
\usepackage{siunitx}

\begin{document}

\preprint{APS/123-QED}

\title{Inversion and Tunability of Van Hove Singularities \\ in $A$V$_{3}$Sb$_{5}$ ($A$ = K, Rb, and Cs) kagome metals}

\author{Sangjun Sim}
\affiliation{
 Department of Physics, Korea Advanced Institute of Science and Technology (KAIST), Daejeon 34141, Korea
}
\author{Min Yong Jeong}
\affiliation{
 Department of Physics, Korea Advanced Institute of Science and Technology (KAIST), Daejeon 34141, Korea
}
\author{Hyunggeun Lee}
\affiliation{
 Department of Physics, Korea Advanced Institute of Science and Technology (KAIST), Daejeon 34141, Korea
}\textbf{}
\author{Dong Hyun David Lee}
\affiliation{
 Department of Physics, Korea Advanced Institute of Science and Technology (KAIST), Daejeon 34141, Korea
}
\author{Myung Joon Han}
 \email{mj.han@kaist.ac.kr}
\affiliation{
 Department of Physics, Korea Advanced Institute of Science and Technology (KAIST), Daejeon 34141, Korea
}

\date{\today}

\begin{abstract}
To understand the alkali-metal-dependent material properties of recently discovered $A$V$_{3}$Sb$_{5}$ ($A$ = K, Rb, and Cs), we conducted a detailed electronic structure analysis based on first-principles density functional theory calculations. Contrary to the case of $A$ = K and Rb, the energetic positions of the low-lying Van Hove singularities are reversed in CsV$_{3}$Sb$_{5}$, and the characteristic higher-order Van Hove point gets closer to the Fermi level. We found that this notable difference can be attributed to the chemical effect, apart from structural differences. Due to their different orbital compositions, Van Hove points show qualitatively different responses to the structure changes. A previously unnoticed highest lying point can be lowered, locating close to or even below the other ones in response to a reasonable range of bi- and uni-axial strain. Our results can be useful in better understanding the material-dependent features reported in this family and in realizing experimental control of exotic quantum phases.
\end{abstract}

\maketitle

\section{Introduction}

Recently discovered superconducting kagome metals $A$V$_{3}$Sb$_{5}$ ($A$ = K, Rb, and Cs) have attracted wide attention because they host multiple quantum phases intertwined with each other, including charge density wave (CDW), superconductivity, and nematic order \cite{ortiz_new_2019, ortiz_csv3sb5_2020, ortiz_superconductivity_2021,zhao_cascade_2021,li_rotation_2022, nie_charge-density-wave-driven_2022, xu_three-state_2022,li_unidirectional_2023,zhong_testing_2023,zhong_nodeless_2023,luo_unique_2023}. The CDW instability is commonly observed in the temperature range below $T_{\text{CDW}} \sim$ 78--102K. Below $T_{\text{CDW}}$, other types of unconventional breakings of time-reversal \cite{jiangUnconventionalChiralCharge2021a,shumiya_intrinsic_2021, wang_electronic_2021,mielke_time-reversal_2022,khasanov_time-reversal_2022,guo_switchable_2022} and rotational symmetry \cite{li_rotation_2022, nie_charge-density-wave-driven_2022, xu_three-state_2022,sur_optimized_2023,wu_unidirectional_2023,asaba_evidence_2024,he_anharmonic_2024} have been reported as well. The superconducting phase is stabilized at $T_{\rm SC}\sim$ 1--3K \cite{ortiz_new_2019, ortiz_superconductivity_2021}.

While these features are common to these metals with some degree of quantitative differences, distinctive features or material dependence have also been clearly observed. For example, scanning tunneling microscopy (STM) studies have reported a $4\times 1$ charge order in CsV$_{3}$Sb$_{5}$ \cite{zhao_cascade_2021,PhysRevX.13.031030,li_unidirectional_2023} and RbV$_{3}$Sb$_{5}$ \cite{shumiya_intrinsic_2021,yu_evolution_2022,meng_manipulating_2023}, whereas it is not observed in KV$_{3}$Sb$_{5}$ \cite{jiangUnconventionalChiralCharge2021a,li_rotation_2022,li_unidirectional_2023}. The CDW modulation pattern along the $c$-axis direction varies depending on the type of alkali metal. To the best of our knowledge, the additional $4c_{0}$ modulation has only been reported in CsV$_{3}$Sb$_{5}$ \cite{ortiz_fermi_2021,stahl_temperature-driven_2022, kautzsch_structural_2023,xiao_coexistence_2023,PhysRevB.108.115123}. Discrepancies were also noted between recent nuclear quadrupole resonance (NQR) \cite{mu_tri-hexagonal_2022, PhysRevResearch.5.L012017} and angle-resolved photoemission spectroscopy (ARPES) experiments \cite{PhysRevResearch.4.033072,kang_charge_2022,hu_coexistence_2022,jiang_observation}. The superconducting phase diagram as a function of doping appears to be different as well: A `second dome' above 35\% Sn doping was reported for $A$ = Cs \cite{oey_fermi_2022, kang_charge_2022}, whereas KV$_{3}$Sb$_{5}$ and RbV$_{3}$Sb$_{5}$ seem to exhibit a single-dome superconductivity \cite{oey_fermi_2022, oey_tuning_2022, kang_charge_2022}. A qualitative difference in superconducting gap symmetry was also noticed. Guguchia \textit{et al}. \cite{guguchia_tunable_2023} recently reported that KV$_{3}$Sb$_{5}$ and RbV$_{3}$Sb$_{5}$ undergo the nodal-to-nodeless gap transition as a function of pressure, but this was not observed for $A$ = Cs \cite{gupta_microscopic_2022,gupta_two_2022, zhang_nodeless_2023}. In the pressure phase diagram, $T_{\rm SC}$ approaches zero below 20--28.8 GPa for KV$_{3}$Sb$_{5}$ and RbV$_{3}$Sb$_{5}$ \cite{zhu_double-dome_2022}. According to Yu \textit{et al.} \cite{yu_pressure-induced_2022}, on the other hand, the second superconducting dome for CsV$_{3}$Sb$_{5}$ persists up to a hydrostatic pressure of 150 GPa.

For a clearer understanding of this intriguing kagome metal family, it is therefore important to have detailed material-specific information as well as further experimental clarification of the controversial issues. Here, we note the principal roles of the Van Hove singularity (VHS) on which many previous theoretical and experimental studies focused \cite{yu_chiral_2012,kiesel_sublattice_2012,kiesel_unconventional_2013,wang_competing_2013,wu_nature_2021,kang_twofold_2022,hu_rich_2022,dong_loop-current_2023, PhysRevB.107.184504,wu_sublattice_2023}. As an important intrinsic characteristic of the kagome lattice band, VHS is considered to be the main source of the wide variety of cascading instabilities observed in this system. At least there are three VHSs have been identified in the vicinity of the Fermi level ($E_F$) \cite{labollita_tuning_2021,consiglio_van_2022,jeong_crucial_2022}. Apparently, several different phases can emerge out of VHSs in combination with electronic correlation \cite{yu_chiral_2012,kiesel_sublattice_2012, kiesel_unconventional_2013, wang_competing_2013, wu_nature_2021, dong_loop-current_2023, PhysRevB.107.184504,wu_sublattice_2023}. Furthermore, a detailed electronic structure analysis has revealed the intriguing nature of VHSs, such as the unexpectedly important role of the out-of-plane Sb-$p$ state and the different energy sequences depending on $A$ \cite{labollita_tuning_2021,jeong_crucial_2022,ritz_impact_2023, li_origin_2023, ritz_superconductivity_2023}. Still, this is far from being a clear understanding of the material's specific characters.

In this paper, we perform first-principles electronic structure calculations based on density functional theory (DFT) together with a tight-binding analysis. In particular, we focus on the different VHS structures in three different materials of $A$ = K, Rb, and Cs. The energetic position and the orbital character of the four VHSs are examined in a comparative manner. The inverted sequence of two low-lying VHSs (one of which is known to have a high-order nature and can possibly enhance the rotational symmetry-broken charge order \cite{kang_twofold_2022, hu_rich_2022,PhysRevB.107.184504}) is found only in CsV$_{3}$Sb$_{5}$. This notable difference is attributed solely to chemistry rather than the structural difference. Because of the different orbital compositions, the four VHSs show qualitatively different responses to uni- and bi-axial strain. The highest-lying VHS (named VHS4), which was not previously noticed, can be lowered enough to be comparable with other VHSs. Our results provide useful information to help understand the intriguing $A$-cation dependence observed in this family and to explore the possible control of their quantum phases.

\section{Computational details}

We carried out first-principles DFT calculations using projector augmented-wave (PAW) method \cite{PhysRevB.50.17953} in the Vienna $ab$ $initio$ simulation package (VASP) \cite{kresse_efficiency_1996_1,kresse_efficient_1996_2,kresse_ultrasoft_1999}. We used the generalized gradient approximation (GGA) as parameterized by Perdew, Burke, and Ernzerhof (PBE) for the exchange-correlation functional \cite{perdew_generalized_1996} and `Grimme's DFT-D3' method in the zero-damping variant \cite{grimme_consistent_2010} for the van der Waals correction. Both lattice parameters and internal coordinates were optimized until the residual forces became less than 1 meV/$\text{\AA}$. For further electronic analysis, we also used the OpenMX \cite{ozaki_variationally_2003,openmx} package. For OpenMX pseudo-atomic basis orbitals, we employed $s$2$p$2$d$1 for K and Rb, $s$2$p$2$d$1$f$1 for Cs, $s$3$p$2$d$2 for V, and $s$2$p$1$d$2$f$1 for Sb. The atomic cutoff radii for the alkali atoms of K, Rb, and Cs were 10.0, 11.0, and 12.0 Bohr, respectively. The radii for V and Sb were chosen to be 6.0, and 7.0 Bohr, respectively. The 12$\times$12$\times$6 $k$-point mesh and the 400 Ry energy cutoff were used. We used our `DFTforge' \cite{yoon_jx_2020} code to analyze the detailed electronic structure further. The charge analysis was conducted with the Löwdin transformation \cite{10.1063/1.1747632}. Tight-binding analysis was performed by constructing maximally localized Wannier functions using the Wannier90 interfaced with OpenMX \cite{PhysRevB.56.12847, PhysRevB.65.035109}. In addition, the calculation results were double-checked using the `HSE(Heyd-Scuseria-Ernzerhof)-06' exchange-correlation functional \cite{krukau_influence_2006}. For this calculation, 6 $\times$ 6 $\times$ 4 $\Gamma$-centered $k$-points and the energy cutoff of 500 eV were used. The energy criterion was set to 10$^{-7}$eV. As GGA gives rise to the better agreement with ARPES data, we present it as our main result.

% Fig.1
\begin{figure}
	\centering
	\includegraphics[width=0.9\linewidth]{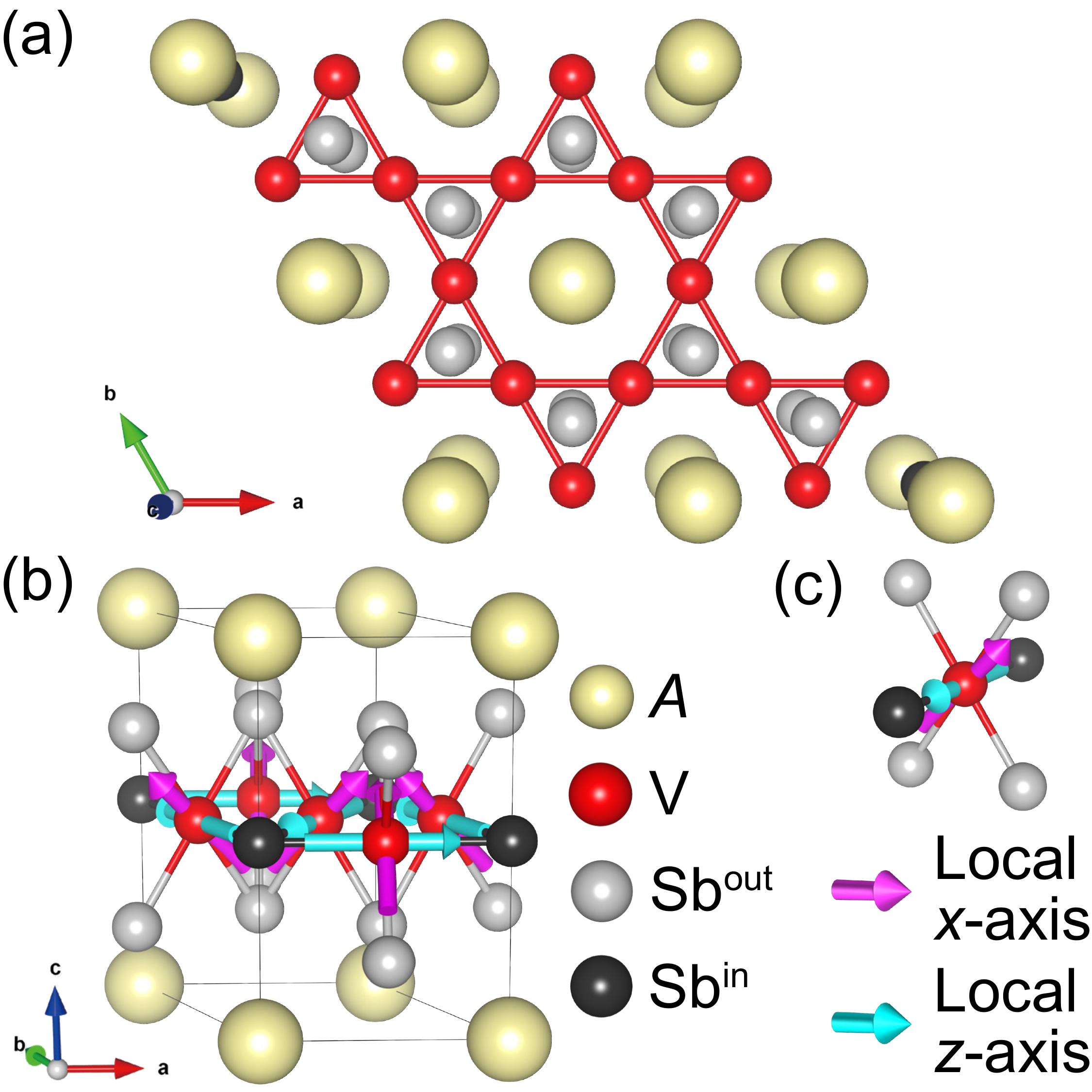}
	\caption{(a) Top view of kagome metal $A$V$_{3}$Sb$_{5}$ ($A$ = K, Rb, and Cs). The yellow, red, gray, and black spheres represent $A$, V, Sb$^{\text{out}}$, and Sb$^{\text{in}}$ atoms, respectively. The connecting red line shows the V-kagome lattice. (b) Three-dimensional view of the $A$V$_{3}$Sb$_{5}$ crystal structure. The magenta- and cyan-colored arrows indicate the local $x$- and $z$-axes, respectively. (c) The local structure of the VSb$_{6}$ octahedron where six Sb atoms (Sb$^{\text{out}}$ and Sb$^{\text{in}}$) surround a V atom. The magenta- and cyan-colored arrows show the same local axes.}
	\label{fig:image-1}
\end{figure}

\section{Results and discussion}

The crystal structure of $A$V$_{3}$Sb$_{5}$ belongs to the $P6/mmm$ space group (No. 191). As shown in Fig.~\ref{fig:image-1}, a V$_{3}$Sb$^{\text{in}}$ layer contains a V-kagome net with Sb$^{\text{in}}$ (in-plane Sb atoms) occupying the vacant center positions of the V hexagons. Sb$^{\text{out}}$ (out-of-plane Sb), forming a honeycomb lattice, is located above and below V$_{3}$Sb$^{\text{in}}$. Alkali metals are interposed between V$_{3}$Sb$_{5}$ layers so that the interlayer distance depends on the size of $A$. Our geometry optimization gives rise to $a$ = $b$ = 5.423$\text{\AA}$ and $c$ = 8.886$\text{\AA}$ for K, $a$ = $b$ = 5.437$\text{\AA}$ and $c$ = 9.056$\text{\AA}$ for Rb, and $a$ = $b$ = 5.453$\text{\AA}$ and $c$ = 9.308$\text{\AA}$ for Cs. These values are in good agreement with experiments \cite{ortiz_new_2019} as well as previous calculations \cite{labollita_tuning_2021, jeong_crucial_2022, consiglio_van_2022}.

% Fig.2
\begin{figure*}
	\centering
	\includegraphics[width=0.93\linewidth]{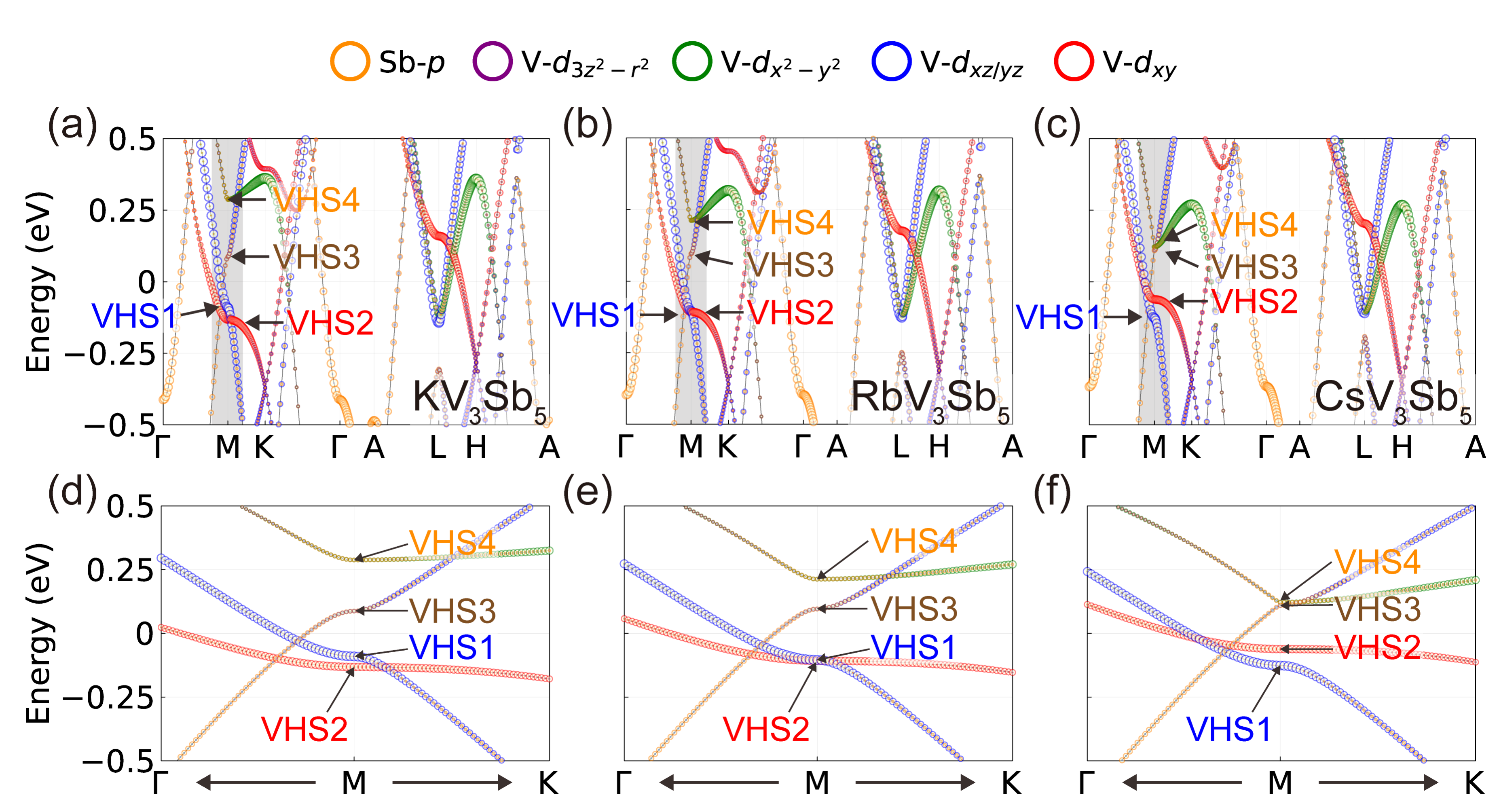}
	\caption{(a--g) The orbital-projected band structures of (a) and (d) KV$_{3}$Sb$_{5}$, (b) and (e) RbV$_{3}$Sb$_{5}$, and (c) and (f) CsV$_{3}$Sb$_{5}$. The shaded region around M in (a)--(c) is highlighted (d)--(f). The Fermi level is set to zero. The dark orange, purple, green, blue, and red colors represent the orbital characters of Sb-$p$, V-$d_{3z^{2}\text{-}r^{2}}$, V-$d_{x^{2}\text{-}y^{2}}$, V-$d_{xz\text{/}yz}$, and V-$d_{xy}$, sequentially. The orbital projection was performed based on the local coordinates for all three kagome metals as shown in Fig.~\ref{fig:image-1}(c). The arrows indicate the location of four Van Hove singularities at M point. The labeling of each VHS follows the energetic sequence in CsV$_{3}$Sb$_{5}$.}
	\label{fig:image-2}
\end{figure*}

\subsection{Electronic structure}

The important common features of the three alkali metal variants in terms of their electronic structure include the electron-like Fermi pocket at $\Gamma$, multiple VHSs at M, and Dirac crossing at K point, which has been scrutinized in a previous first-principles calculation and angle-resolved photoemission spectroscopy (ARPES) \cite{labollita_tuning_2021,cho_emergence_2021,kang_twofold_2022,luo_electronic_2022,hu_rich_2022,consiglio_van_2022,jeong_crucial_2022}; see Figs.~\ref{fig:image-2}(a--c). Of particular interest among them are the VHSs, as they are deemed to be responsible for the multiple ordered phases cascading with lowering temperature \cite{yu_chiral_2012,kiesel_sublattice_2012, wang_competing_2013, kiesel_unconventional_2013,wu_nature_2021}. For convenience, we took CsV$_{3}$Sb$_{5}$ as a reference and indexed the four VHSs in the increasing order of their energy; VHS1, VHS2, VHS3, and VHS4 (see Fig.~\ref{fig:image-2}(c)). As shown in Figs.~\ref{fig:image-2}(a--c), VHS1 and VHS2 are commonly located below $E_F$, whereas VHS3 and VHS4 are above. In the literature, the highest-lying VHS4 has been on the periphery of attention, as it is located far away from $E_F$.

To analyze the orbital characters of the VHSs, we chose a local axis coordinate considering the geometry of the VSb$_6$ octahedron \cite{jeong_crucial_2022}; see Fig.~\ref{fig:image-1}(c). Each orbital component is represented by different colors in Fig.~\ref{fig:image-2}. VHS1 and VHS2 are composed mainly of V-$d_{xz\text{/}yz}$ and V-$d_{xy}$, respectively, with their weights more than 60$\%$ for all three cases of K, Rb, and Cs (see Table~\ref{tab:table-1}). The contribution from Sb$^{\rm out}$-$p$ at VHS1 and VHS2 is relatively small; less than 5\%. For VHS3 and VHS4, on the other hand, the Sb$^{\rm out}$-$p$ contribution is sizable, as partly discussed in a previous study of $A$ = Cs \cite{jeong_crucial_2022}. For $A$ = Cs, the Sb$^{\rm out}$-$p$ portion is about 30\% at VHS3 and VHS4. For $A$ = K (Rb), it is 26\% (28\%) and 23\% (30\%) at VHS3 and VHS4, respectively.

Another important point is the relative position of VHS1 and VHS2 depending on $A$. Let us first note that, in $A=$ K and Rb (Figs.~\ref{fig:image-2}(a) and (b)), VHS2 is lower in energy than VHS1, whereas their positions are reversed in CsV$_{3}$Sb$_{5}$. This was noted by Labollita and Botana \cite{labollita_tuning_2021}. In fact, VHS2 gradually moves upward from $-$132 meV for K to $-$108 and $-$62 meV for Rb and Cs. On the contrary, VHS1 shows a decreasing trend; $E_{\rm VHS1}=$ $-$90, $-$104, and $-$126 meV for K, Rb, and Cs, respectively. In KV$_{3}$Sb$_{5}$ and RbV$_{3}$Sb$_{5}$, $E_{\rm VHS2}$ is lower than $E_{\rm VHS1}$, and as a result, VHS1 is located closer to $E_F$, which is not the case for Cs.

It is noted that the `higher-order' nature of VHS2 has been highlighted in previous studies \cite{kang_twofold_2022,chandrasekaran_effect_2022,hu_rich_2022,jeong_crucial_2022,chandrasekaran_practical_2023}. The faster divergence of its density of states and the weaker Fermi surface nesting were presumed to be important for, e.g., nematic charge order and nodal superconductivity \cite{PhysRevLett.123.207202,yuan_magic_2019,PhysRevResearch.1.033206, PhysRevB.107.184504}. Thus, the closer location in CsV$_{3}$Sb$_{5}$ to the Fermi energy can possibly be related to the observed difference in characteristics from those of KV$_{3}$Sb$_{5}$ and RbV$_{3}$Sb$_{5}$.

\subsection{Chemical effect}

\begin{figure}
	\centering
	\includegraphics[width=0.9\linewidth]{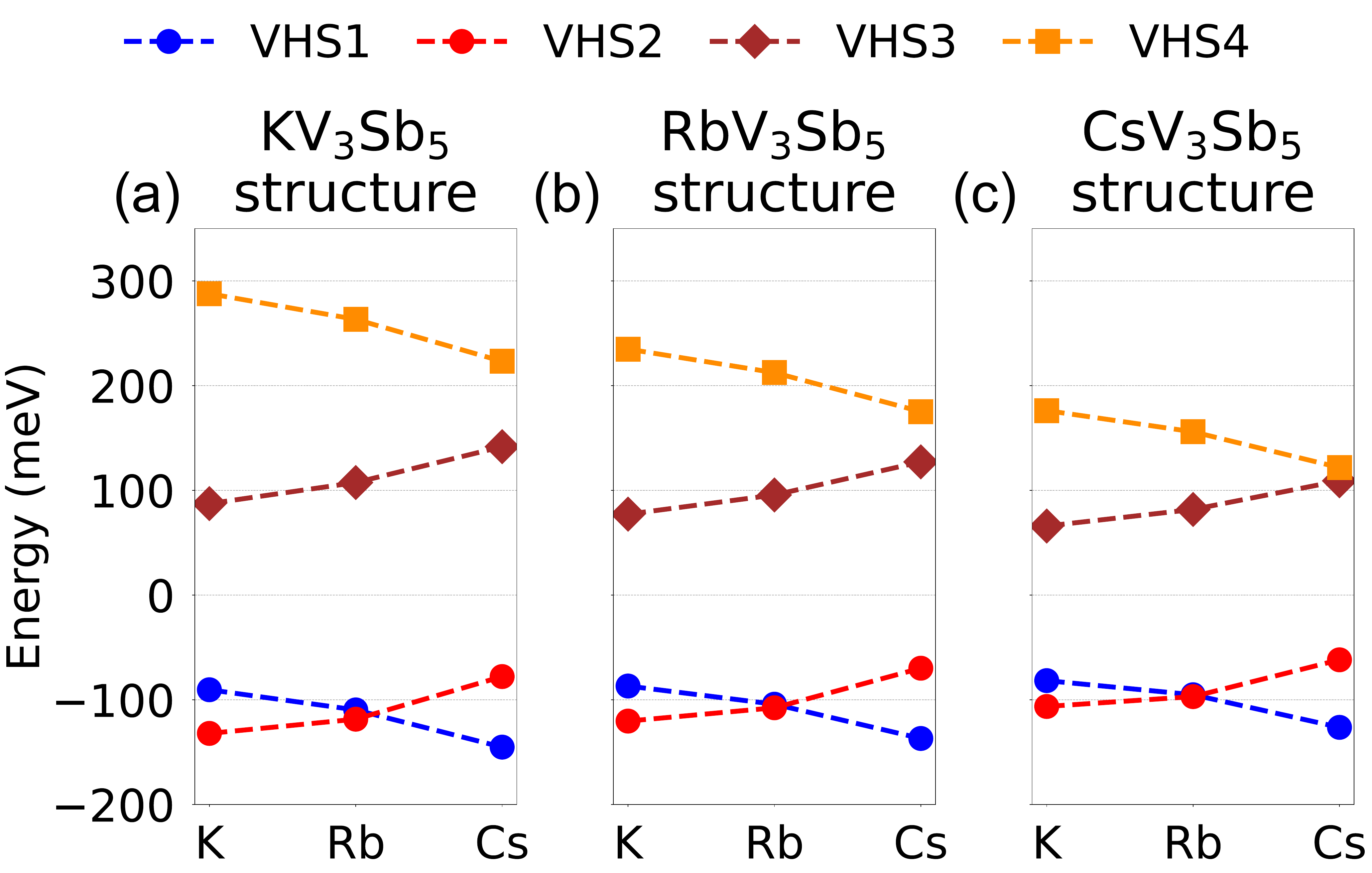}
	\caption{(a-c) The calculated VHS positions in the fixed crystal structures. The optimized (a) KV$_{3}$Sb$_{5}$, (b) RbV$_{3}$Sb$_{5}$, and (c) CsV$_{3}$Sb$_{5}$ structures were used with different $A$ element substitutions. The Fermi level is set to zero for all cases. The blue, red, brown, and orange colors represent VHS1, VHS2, VHS3, and VHS4, sequentially.
	}
	\label{fig:image-3}
\end{figure}

To elucidate the underlying origin of this intriguing material dependence of VHSs, it is important to examine the effects coming from chemistry and structure. Figure~\ref{fig:image-3} summarizes the calculated results of VHS positions with the fixed atomic structures. Figures~\ref{fig:image-3}(a--c) correspond to the structures optimized with $A$ = K, Rb, and Cs, respectively. In each structure, we performed three different calculations, with the insertion of $A$ = K, Rb, and Cs. It was noted that VHS1 moves downward in its energy as the alkali ion gets heavier (with the fixed geometry; see blue lines), whereas VHS2 goes upward (red lines). In the case of $A$ = K structure (Fig.~\ref{fig:image-3}(a)), for instance, VHS1 was found at $-$90, $-$110, and $-$145 meV with $A$-site substitution of K, Rb, and Cs, respectively. VHS2 was at $-$132, $-$119, and $-$78 meV for K, Rb, and Cs, respectively. This feature is commonly observed in all three structural cases, as presented in Figs.~\ref{fig:image-3}(b) and (c). Interestingly, with $A$ = Cs, the relative positions of VHS1 and VHS2 were always found to be reversed; namely, $E_{\rm VHS1} < E_{\rm VHS2}$ regardless of the structure. It is, therefore, solely attributed to the `chemical effect' of the alkali metals on the energetics of the VHSs \cite{PhysRevB.82.134408}.

This finding was partly supported by our tight-binding analysis. Our parameterization based on maximally localized Wannier functions gives rise to the $A$-site $s$-orbital on-site energies of $-$1.325, $-$1.209, and $-$1.003 eV, respectively, for KV$_{3}$Sb$_{5}$, RbV$_{3}$Sb$_{5}$, and CsV$_{3}$Sb$_{5}$. This is consistent with the increasing trend observed for VHS2, which has the non-negligible portion of $A$-$s$ contribution (see Table~\ref{tab:table-1}). The same was also true for VHS3 and $A$-$p$ whose on-site energies were found to be 0.301, 0.537, and 0.758 eV for $A$ = K, Rb, and Cs, respectively. Although the chemical effect of $A$ cation cannot wholly be captured by two tight-binding parameters, and the real band dispersion is quite complicated near $E_F$, it provides useful information.

Given that VHS(s) in the vicinity of $E_F$ is the key to inducing multiple quantum phases in this family \cite{PhysRevLett.123.207202,yuan_magic_2019,PhysRevResearch.1.033206, PhysRevB.107.184504}, the characteristic features of CsV$_3$Sb$_5$ distinctive from KV$_3$Sb$_5$ and RbV$_3$Sb$_5$ can possibly be related to its VHS energetics. It is also noteworthy that VHS2 is the only one presumed to be higher-order in nature \cite{kang_twofold_2022,hu_rich_2022}. A detailed and systematic further study focusing on this relation could be a useful future direction.

% Fig.3

% Table for orbital occupation
\begin{table*}
%	\footnotesize
	\centering
	\caption{The calculated orbital portions of the four VHSs in $A$V$_{3}$Sb$_{5}$ ($A$ = K, Rb, and Cs). The orbital projections were performed with atomic radii of $R_{\rm V} = 6.0$, $R_{\rm Sb} = 7.0$, $R_{\rm K} = 10.0$, $R_{\rm Rb} = 11.0$, and $R_{\rm Cs} = 12.0$ Bohr. The other orbital portions not shown here are approximately 30--35\%.}
	\label{tab:table-1}
	\begin{tabular}{ c | c c c c | c c c c | c c c c | c c c c}
		\hline
		\multirow{2}{*}{Unit (\%)} &
		\multicolumn{4}{c}{$A$-$s$} &
		\multicolumn{4}{c}{V-$d_{xy}$} &
		\multicolumn{4}{c}{V-$d_{x^2-y^2}$} &
		\multicolumn{4}{c}{Sb$^{\rm in}$-$p$} \\ \cline{2-17}
		& VHS1 & VHS2 & VHS3 & VHS4 & VHS1 & VHS2 & VHS3 & VHS4 & VHS1 & VHS2 & VHS3 & VHS4 & VHS1 & VHS2 & VHS3 & VHS4 \\
		\hline
		KV$_{3}$Sb$_{5}$ & 0.00 & 3.21 & 0.00 & 0.00 & 0.00 & 64.72 & 0.00 & 0.00 & 0.00 & 0.00 & 1.31 & 31.88 & 0.00 & 0.00 & 4.96 & 0.00 \\
		\hline
		RbV$_{3}$Sb$_{5}$ & 0.00 & 2.84 & 0.00 & 0.00 & 0.00 & 64.49 & 0.00 & 0.00 & 0.00 & 0.00 & 0.77 & 32.82 & 0.00 & 0.00 & 5.45 & 0.00 \\
		\hline
		CsV$_{3}$Sb$_{5}$ & 0.00 & 2.50 & 0.00 & 0.00 & 0.00 & 64.05 & 0.00 & 0.00 & 0.00 & 0.00 & 0.26 & 34.62 & 0.00 & 0.00 & 6.19 & 0.00 \\
		\hline
		\multirow{2}{*}{Unit (\%)} &
		\multicolumn{4}{c}{$A$-$p$} &
		\multicolumn{4}{c}{V-$d_{xz/yz}$} &
		\multicolumn{4}{c}{V-$d_{3z^2-r^2}$} &
		\multicolumn{4}{c}{Sb$^{\rm out}$-$p$} \\ \cline{2-17}
		& VHS1 & VHS2 & VHS3 & VHS4 & VHS1 & VHS2 & VHS3 & VHS4 & VHS1 & VHS2 & VHS3 & VHS4 & VHS1 & VHS2 & VHS3 & VHS4 \\
		\hline
		KV$_{3}$Sb$_{5}$ & 0.00 & 0.00 & 1.13 & 0.00 & 74.80 & 0.01 & 30.22 & 0.00 & 0.00 & 0.92 & 0.00 & 0.00 & 5.07 & 3.77 & 26.69 & 30.46 \\
		\hline
		RbV$_{3}$Sb$_{5}$ & 0.00 & 0.00 & 1.09 & 0.00 & 74.77 & 0.01 & 29.31 & 0.00 & 0.00 & 1.23 & 0.00 & 0.00 & 4.89 & 4.25 & 27.99 & 30.77 \\
		\hline
		CsV$_{3}$Sb$_{5}$ & 0.00 & 0.00 & 1.82 & 0.00 & 74.59 & 0.01 & 28.17 & 0.00 & 0.00 & 1.95 & 0.00 & 0.00 & 4.89 & 5.42 & 28.94 & 29.64 \\
		\hline
	\end{tabular}
\end{table*}

\subsection{Structural effect}

The structure effect on the VHS positions can also be seen in Figs.~\ref{fig:image-3}(a--c). Let us consider the case of $A$ = K, for example. The energy location of VHS2 becomes higher in the optimized structure with a heavier alkali-metal; $E_{\rm VHS2}$ = $-$132, $-$120, and $-$106 meV in the optimized structure of KV$_{3}$Sb$_{5}$, RbV$_{3}$Sb$_{5}$, and CsV$_{3}$Sb$_{5}$, respectively. The increasing feature is also observed in VHS1. On the other hand, the behavior of VHS3 and VHS4 is the opposite; namely, their energies get lower in the geometry optimized with heavier alkali atoms. The downward shifting of VHS4 is significantly greater, and indeed, the Sb$^{\rm out}$-$p$ portion is greatest in VHS4. For detailed quantitative information on the orbital components in each VHS, see Table~\ref{tab:table-1}. Further analysis based on maximally localized Wannier functions also showed that the tight-binding hopping parameter between V-$d_{x^{2}\text{-}y^{2}}$ and Sb$^{\rm out}$-$p$ exhibits sizable dependence on the corresponding structural change.

% Fig.4
\begin{figure}
	\centering
	\includegraphics[width=0.9\linewidth]{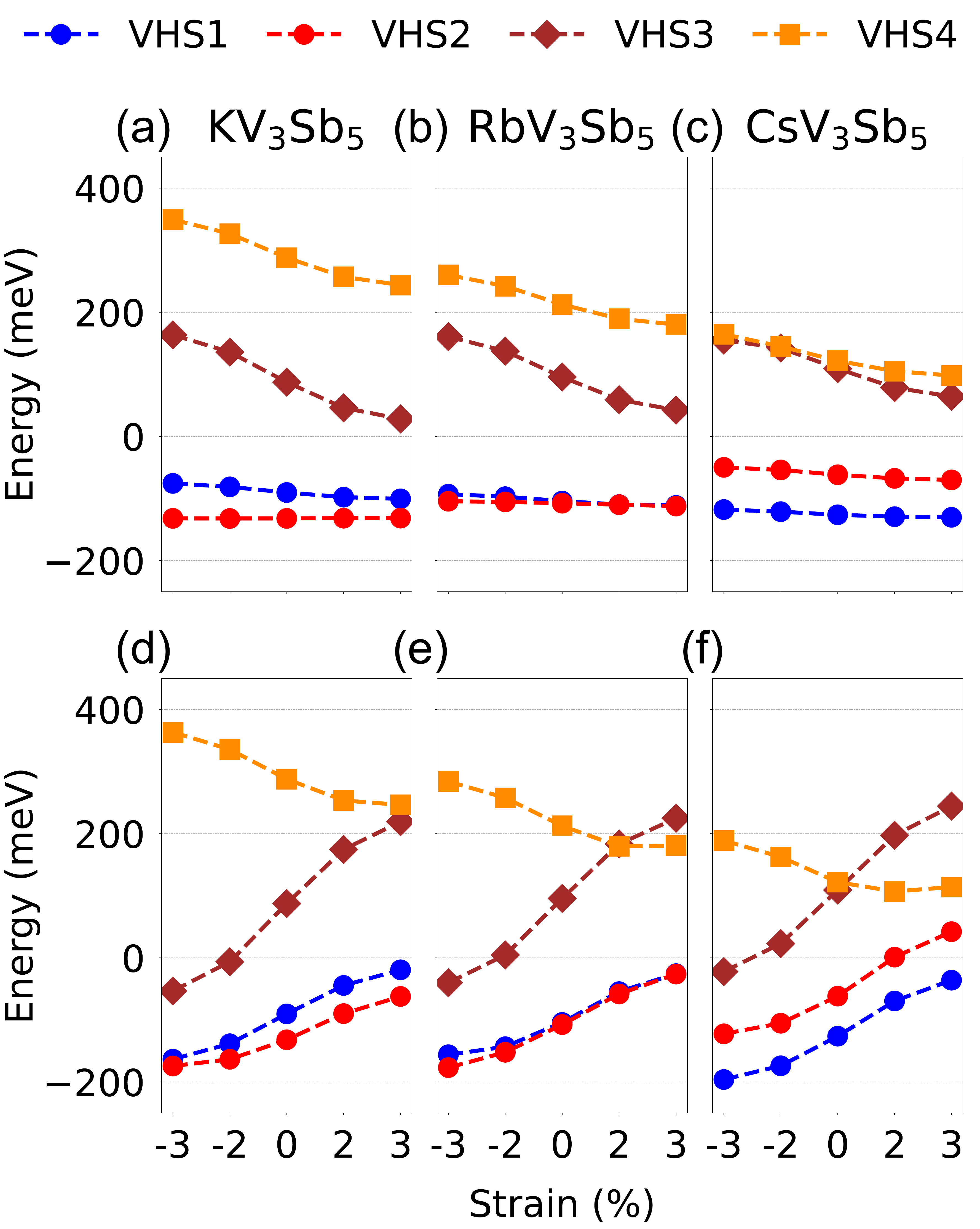}
	\caption{(a--c) The energetic positions of four VHSs in $A$V$_{3}$Sb$_{5}$ ($A$ = K, Rb, and Cs) in response to compressive (negative sign) and tensile (positive sign) strain up to $\pm$3\% along $c$-axis. (d--f) The calculated positions of the VHSs under biaxial compressive and tensile strains along the $a$-$b$ axis. The blue, red, brown, and dark orange colors represent VHS1, VHS2, VHS3, and VHS4, respectively. The Fermi energy is set to be zero.}
	\label{fig:image-4}
\end{figure}

\subsection{Tuning the energy levels of VHSs by strain}

The above result motivated us to investigate the possible use of strain to control the VHS positions. Figures~\ref{fig:image-4}(a--c) show the calculated VHSs in response to the $c$-lattice parameter change. While this uniaxial strain effect is relatively weak for VHS1 and VHS2, VHS3 and VHS4 were significantly affected, and both gradually shifted down toward $E_F$ as the $c$ lattice parameter increased. While this result is consistent with a previous study by Consiglio {\it et al}. \cite{consiglio_van_2022}, our analysis provides additional information and insight, especially for VHS4. Taking KV$_{3}$Sb$_{5}$ as an example, VHS4 is located at +288 meV at zero strain, and it decreases to $\sim$ +244 meV at +3\% tensile strain (see Fig.~\ref{fig:image-4}(a)). On the other hand, VHS1 and VHS2 are originally located at $-$90 meV and $-$132 meV, respectively, and they move to $-$101 meV and $-$132 meV at +3\% strain. 
Once again, this notably different response between the two lower (VHS1 and VHS2) and higher energy VHSs (VHS3 and VHS4) is attributed to the different orbital characters involved in their formation; see Table~\ref{tab:table-1}. Further control may be feasible by employing biaxial strain. Figs.~\ref{fig:image-4}(d--f) shows that a qualitatively different response in VHS positions can be obtained with biaxial strain. From compressive to tensile strain (along the $a$ and $b$-axis), VHS1, VHS2, and VHS3 moved toward the higher energy, whereas the VHS4 position got lower, just as in the uniaxial case. The response of VHS3 is so sizable that its relative position to VHS4 is eventually reversed under a large tensile strain limit in RbV$_{3}$Sb$_{5}$ and CsV$_{3}$Sb$_{5}$; see Figs.~\ref{fig:image-4}(e) and (f).

\section{Conclusions}

In summary, we investigated the alkali-metal-dependent electronic structure of $A$V$_3$Sb$_5$ ($A$ = K, Rb, and Cs) to understand the previously reported intriguing material dependence. Our calculation results show that the energetic positions of low-lying Van Hove singularities in CsV$_{3}$Sb$_{5}$ are distinctive from those of $A$ = K and Rb, and the characteristic higher-order Van Hove point is located closer to $E_F$. Detailed electronic analyses revealed that this notable difference can be attributed to chemical effects apart from structures. Additionally, the different orbital compositions lead the four VHSs to exhibit qualitatively different responses to strain. In particular, the highest-lying VHS4 can be lowered enough to be comparable with other VHSs. This distinct response of VHS4 to strain demonstrates the useful potential for strain engineering of these materials. Our findings provide valuable insights into understanding the $A$-cation dependence observed in experiments and shed new light on understanding recent experiments and approaches to controlling the intertwined quantum phases in this material family.

\section{Acknowledgements}
We thank Sergey Savrasov for the useful discussion. This work was supported by the National Research Foundation of Korea (NRF) grant funded by the Korea government (MSIT) (Grant Nos. 2021R1A2C1009303 and RS-2023-00253716).

\nocite{*}

\end{document}